 \documentclass[final,5p,twocolumn]{elsarticle}

\usepackage{amssymb}

\usepackage{wrapfig}

\journal{arXiv}

\begin{document}

\begin{frontmatter}


%
\title{Entropic analysis of a hierarchically organized Axelrod model.}

\author[1,2]{Marcos E. Gaudiano}
\ead{marcosgaudiano@gmail.com}
\author[3,2]{Jorge A. Revelli}

\address[1]{Centro de Investigaci\'on y Estudios de Matem\'atica. Consejo Nacional de Investigaciones Cient\'\i ficas y T\'ecnicas.}
\address[3]{Instituto de F\'\i sica Enrique Gaviola. Consejo Nacional de Investigaciones Cient\'\i ficas y T\'ecnicas.}
\address[2]{Universidad Nacional de C\'ordoba. Facultad de Matem\'atica, Astronom\'\i a, F\'\i sica y Computaci\'on.\\
\vspace{0.5cm}
Av. Medina Allende s/n , Ciudad Universitaria, X5000HUA C\'ordoba, Argentina.}

\begin{abstract}
 Hierarchically organized patterns are ubiquitously found in complex systems. However, this point is frequently misrepresented in many Sociophysics models, mainly because random initial conditions are by far the most assumed in the literature.\\  
In this article, we studied a simple and quasi-aparametric Axelrod model of culture dynamics assuming structured (hierarchically organized) initial conditions. As a first remarkable point, the maximum final culture branching is observed to correspond to a highest uncontrollability regime of an entropy based framework described before. 
Also, this model shows a quantization of culture patterns that can be explained with the aid of that formalism.
\end{abstract}


\end{frontmatter}



\begin{section}{Introduction.}
It is a well established fact that both nature and social systems have organized structures \cite{pumain}. Complex system
models often attempt to reproduce those structures. In other words, hierarchically organized structures are usually unveiled through the model features, supposing either random initial configurations or random boundary conditions.\\
However, it should not be the general case. There are many situations where structures determine the evolution of the systems. So hierarchically structured conditions must be taken into account as part of the integral description of the dynamics. This fact implies that structures should have a more predominant role in many models.\\

In recent years, some models incorporated the structure in a novel way.
For instance, Encarna\c{c}\~{a}o {\it et al.} \cite{encarnacao} carried out a 2-D study of Lisbon metropolitan area by using an entropy function, dependent of the built-up area fractal dimension \cite{falconer}. This quantity rise, acceleration, etc. define entropic regimes that allow to study  local hierarchically organized urban layouts. It provides a natural partition of the city consisting of different urban dynamics zones, which would eventually requiry particular planning and/or policies. This methodology was also applied by Sun {\it et al.} \cite{jinsun} for classifying the structure of deforestation areas in the Amazon rainforest. In addition, a 1-D version of this framework was implemented to characterize temporal patterns of public transport strikes \cite{lucca}. Remarkably, the intrinsic meaning of the entropic regimes used in \cite{encarnacao}, \cite{jinsun} and \cite{lucca} turn out to be basically invariant, regardless the fact these works deal with clearly different systems.\\

Actually, the just above commented works can be seen as particular cases of the general framework described in \cite{gaudiano}. In that article, the author explores hierarchically organized patterns in multidimensional complex systems, together with the notion of {\it uncontrollability}. It is provided a classification of system components, which is based on a (generalized) entropy function. We will call it {\it entropic formalism} from now on.\\



So far, we have cited three real world examples. However, the entropic formalism has also interesting applications on widely-known theoretical sociophysics 2-D models, when hierarchically organized initial conditions are considered. For example, Sznajd's model opinion dynamics was studied in \cite{third_position}-\cite{entropico}. One of the main conclusions of these articles is that  even under high political apathy, a third emergent party victory is clearly possible.\\ Another example is the segregation Schelling's model studied in \cite{schelling_entropico}. That article showed how the structure of a social minority clearly determines either the  coexistence with the rest or the ocurrence of segregation.\\
Remarkably, it turns out that system's fate characterization by using the entropic regimes made the three latter works to be minimalistic in terms of parameters. It also allowed the authors to unveil many morals and observable facts (with highly correspondence to reality) that cannot be visualized under random initial  conditions (mostly assumed in the literature).\\

With the above ideas in mind, Axelrod model \cite{axelrod} will be
explored in the present article. It will be an extension of the previous mentioned results to higher dimensional agent models.\\
Though originally created for describing disemination of
cultures, Axelrod model can also be applied to other related issues in opinion dynamics, social mood evolution, consensus through referendums (see e.g. \cite{axelrod_media}, \cite{tangled} and \cite{castellano_fortunato}). We note that our results will be useful for the study of every of those Axelrod model pictures, since this article will basically concern on the initial {\it structured} conditions rather than iteraction rule interpretations.
\\

The article is organized as follows. In Section \ref{the_model}, we introduce Axelrod model under structured initial conditions. 
In Section \ref{convergence_time}, system's response will be analized in terms of convergence time. Uncontrollability notions and the entropic formalism  of \cite{gaudiano} are described in Section \ref{an_entropic_formalism}. Some implications of it are studied in Section \ref{cultures}, which also allowed us to unveil a kind of ideological culture quantization that is explained in Section \ref{quantization}. Conclusions and final remarks are presented in Section \ref{conclusion}.


\end{section}

\begin{section}{The model.}\label{the_model}
Axelrod's model characterizes a set of individuals by their social preferences and ideally describes its time evolution in terms of social inﬂuence and homophily \cite{castellano_fortunato}. In the present article we explore a very simple version of Axelrod dynamics, consisting of an array of $\lambda\times\lambda$ individuals ($\lambda=64$), each one having $F=8$ binary attributes (maybe obtained from a survey). This could provide (e.g.) an essential picture of the mood evolution of a set of people, specially when considering referendums. Additionally, this choice will avoid huge computational efforts associated to the fact that Axelrod's is a {\it vectorial} (3-D) model, essentially harder to simulate than Sznajd's or Schelling's,  which are --as mentioned above-- 2-D agent models.\\

Specifically, the system is mathematically represented by a $\lambda\times\lambda\times F$ tensor $M$, such that $M_{ijk}(t)=\pm 1$ represents an individual located at the site $(i,j)$ whose $k$-th survey answer is $yes/no$ at time $t$, respectively. This defines the agent {\it active} attributes.\\
In addition, the existence of {\it no answer/don't know}'s or simply survey set-up defects is incorporated in the form of information gaps, that do not evolve in time, to which we assign the value $M_{ijk}(t)=0$ at any time $t$. They are agent {\it inactive} attributes which, in the context of dissemination of cultures, could be interpreted as social {\it taboos}. Frequently, we will use both terms, indistinctly.\\
System time evolution is carried out by {\it adjacent} sites. Two sites $(i,j)$ and $(i',j')$ are defined as adjacent if it holds that $0<(i-i')^2+(j-j')^2\leq 2$.\\ 

For a randomly chosen site $(i,j)$ having at least an active attribute, the system will evolve as follows.  With probability
\begin{equation}\label{omega}
\omega=\frac{1}{F}\sum_{k=1}^{F}M_{ijk}(t)M_{i'j'k}(t)\delta_{M_{ijk}(t),M_{i'j'k}(t)},
\end{equation}
being $(i',j')$ an aleatory chosen\footnote{On Eq.(\ref{omega}), $\delta$ means Kronecker's delta} adjacent site of $(i,j)$, it is picked up (if possible) a random $\bar{k}$ such that $M_{ij\bar{k}}(t)$ and $M_{i'j'\bar{k}}(t)$ are both active and distinct attributes. If this can be done, then the update at time $t+\Delta t$ will read
\begin{equation}\label{MijtDeltat}
M_{ij\bar{k}}(t+\Delta t)=M_{i'j'\bar{k}}(t),
\end{equation}
otherwise the system will remain invariant. Note Eqs. (\ref{omega}) and (\ref{MijtDeltat}) do not update the inactive attributes. In turn, every update takes place when both $M_{ij\bar{k}}(t)$ and $M_{i'j'\bar{k}}(t)$ are not zero. The latter type of updates is clearly Axelrod's \cite{castellano_fortunato}.\\

We define as {\it unstable} sites to the agents $(i,j)$ that can be updated through  Eqs. (\ref{omega}) and (\ref{MijtDeltat}). As time goes by, the number of unstable sites simply vanishes (Fig.(\ref{Ni})) and the system converges to an ultimate configuration (Fig.(\ref{T0T})).\\ 

\begin{figure}
\includegraphics[width=0.5\textwidth]{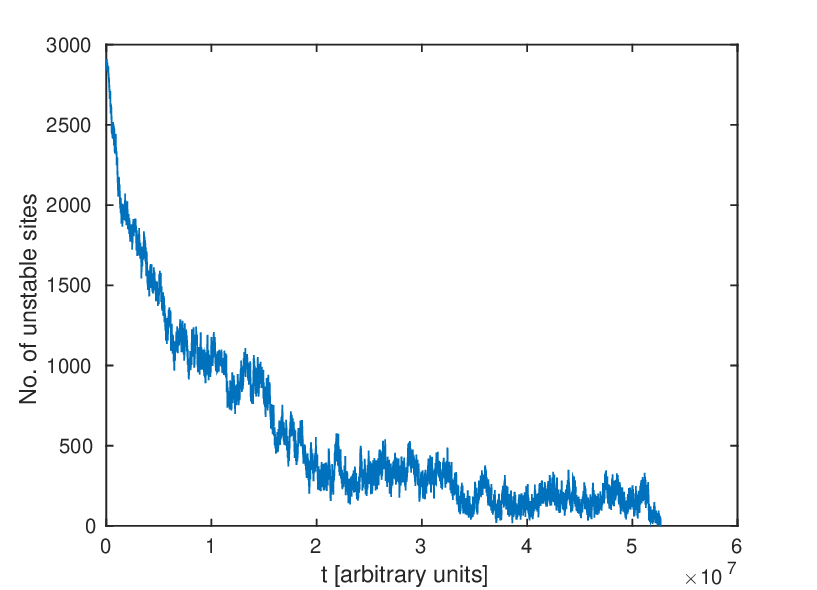}
\caption{Example of time evolution of No. of unstable sites. The initial and ultimate system configurations are shown on Fig.(\ref{T0T}). Note there is no evolution beyond $t\approx 5.27\times 10^7$, in which the No. of unstable sites vanishes.}
\label{Ni}
\end{figure}
\begin{figure*}
\centering
\includegraphics[width=1.1\textwidth]{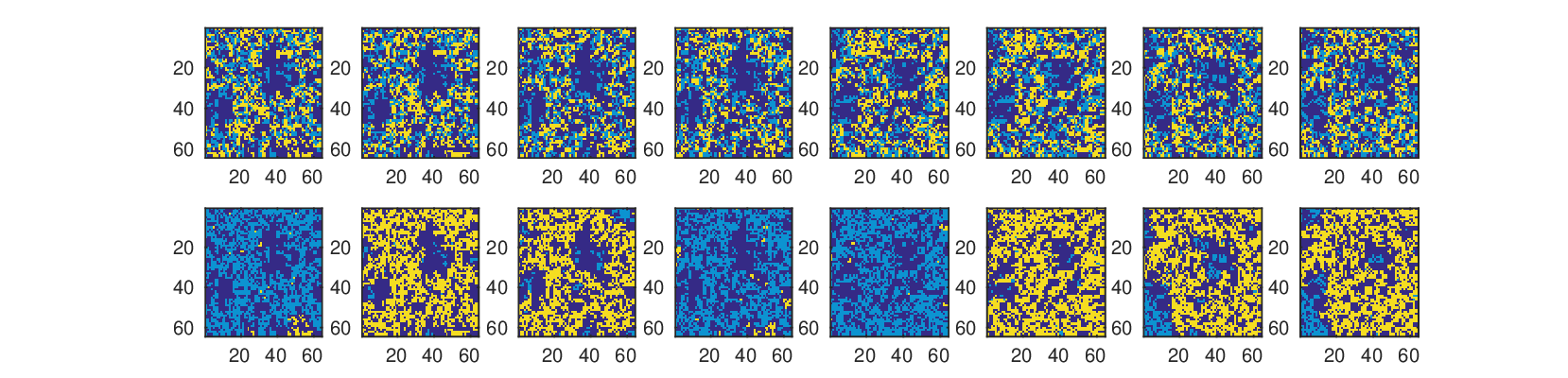}
\caption{Example of system tensor $M$ evolution from $t=0$ (top) to $t=\infty$ (bottom). From left to right, yellow, light blue and blue pixels respectively correspond to $M_{ijk}=+1,-1,0$, while $k=1,...,8$ are the attributes (or survey answers). 50\% of the pixels are agent inactive attributes.  Initially, there is 25$\%$ of each $M=\pm$ answer. There is 21\% (29\%) of light blue (yellow) pixels at $t=\infty$.}
\label{T0T}
\end{figure*}

Due to historic and sociological reasons, real attributes of individuals may show up both geographically and ideologically distributed conforming patterns \cite{morgan}. These kind of arrangements in general cannot be considered random. Consequently, the existence of hierarchically organized structures mentioned above \cite{pumain} will be incorporated as follows.
We assume that individual answer/attributes initially conform $tri-dimensional$ patterns having fractal dimension (computed with the  box-counting algorithm \cite{theiler}). For instance, the initial set of points (or pixels) $(i,j,k)$ with $M_{ijk}(0)= +1 \textrm{(yellow)}, -1 \textrm{(light blue)}$ of Fig.(\ref{T0T}) are not random. They were actually set to have fractal dimension 2.25 and 2.23, respectively (see  {\bf \ref{appendixA}}).\\
Following the colours of the figure, from now on we denote by $Y$ and $L$ to the positive and negative parts of $M_{ijk}(t)$ at time $t$, which will be additionally assumed to have initial fractal dimension $D_Y$ and $D_L$, respectively.
\end{section}

\begin{section}{Convergence time.}\label{convergence_time}
\begin{figure}
\includegraphics[width=0.52\textwidth]{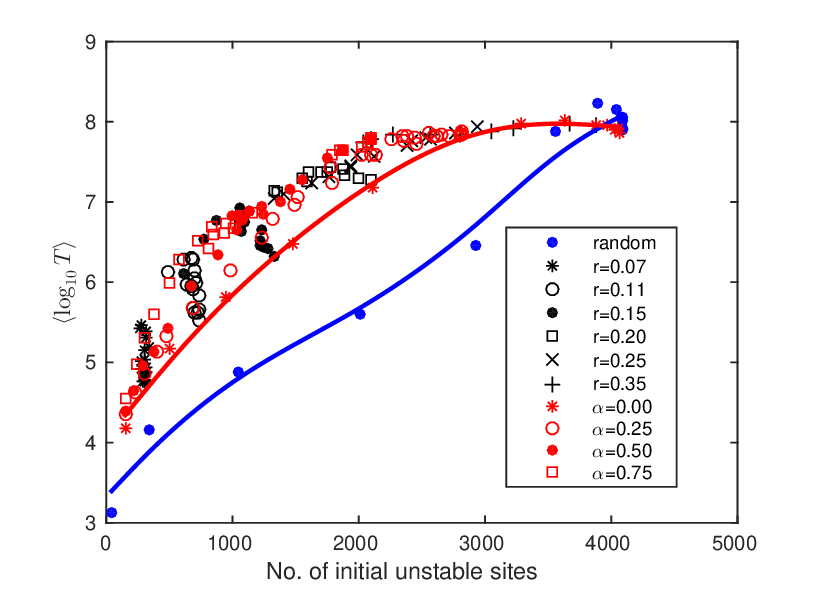}
\caption{As a function of the initial number of unstable sites, structured initial conditions generally show a convergence time that is between 10 and 100 times higher than the random case. The initial fractal $Y$-patterns were generated in two ways: a) varying $\alpha$, for a given $r$; b) varying $r$, for a given $\alpha$. Red and blue curves correspond to the $\alpha=0$ and random cases, respectively. {\it These curved lines are just cubic smoothing spline interpolations of data. They do not represent any model adjustment. The same holds for Figs. (\ref{logt}), (\ref{sigmaDc_DYIII}), (\ref{amarillotabu0}) and (\ref{tabu1_2_012345678}).}}
\label{logt_inestables}
\end{figure}     
\begin{figure}
\includegraphics[width=0.52\textwidth]{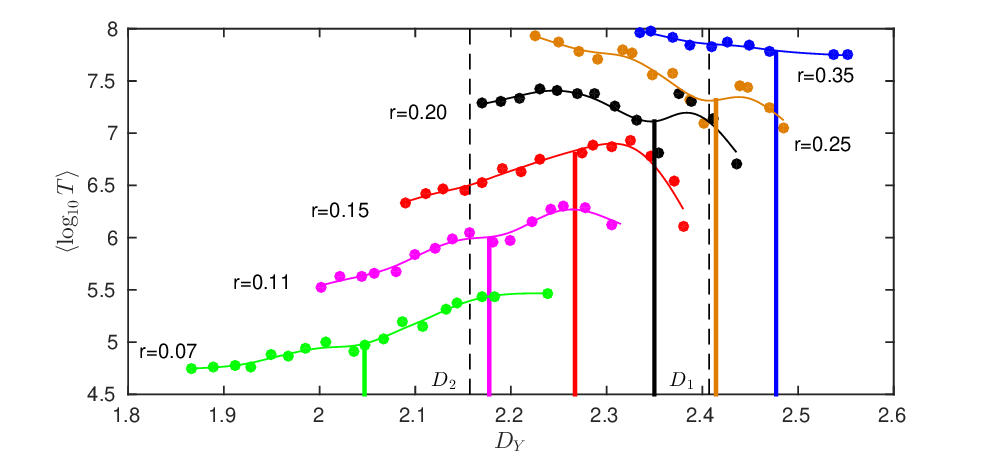}
\caption{The convergence time non-monotically depends on $D_Y$ when the range $D_{min}(r)\leq D_Y\leq D_{max}(r)$ substantially intersects the regime $D_2\leq d \leq D_1$ (Section \ref{an_entropic_formalism}). Solid vertical lines indicate the middle of each range, i.e. $D_Y=\frac{1}{2}(D_{min}(r)+D_{max}(r))$.} 
\label{logt}
\end{figure}
In this section we explore the dependence of the system convergence time $T$ with respect to the initial structured patterns. Numerically, it is observed that every possible pattern of pixels (i.e attributes of the agents) lying inside of the $\lambda\times\lambda\times F$ size box will have a fractal dimension $d$ bounded by
\begin{equation}\label{N262}
N\approx 2.62857... 
\end{equation} 
which is simply the box-counting log fitting slope for a volume of $\lambda\times\lambda\times F$ pixels. Thus, the patterns under consideration can be seen as embedded in a space of non-integer dimensionality.\\
Then, as suggested in \cite{gaudiano}, every volume of $V$ pixels of a pattern with fractal dimension $d$ has necessarily to satisfy:
\begin{equation}\label{LVU}
2^{md}\leq V\leq \lambda^2F(2^m)^{d-N}, \qquad{0\leq d\leq N}
\end{equation}  
where $m=5$ is the No. of steps of the box-counting (or scales involved in the fractal behaviour).\\
 
For the sake of simplicity, we suppose that the patterns of the $L$ and $Y$ parts have initially the same size $r\lambda^2F$, where $0\leq r\leq 0.5$. This way, we will focus on studying how structure can influence the system response.\\
Eq.(\ref{LVU}) points out a lowest fractal dimension $D_{min}(r)$ compatible with that volume determined by $r$. Let us assume that this is the case for the initial $L$-pattern,  i.e. $D_L=D_{min}(r)$ at $t=0$.\\
The dimension of the $Y$-part of the initial patterns will satisfy $D_Y\geq D_L$. It is convenient to introduce here the parameter $\alpha$, which allows to write $D_Y$ as
\begin{equation}
D_Y=(1-\alpha)D_{min}(r)+\alpha D_{max}(r)\qquad{0\leq\alpha\leq 1,}
\end{equation}
where $D_{max}(r)$ is the highest dimension compatible with Eq.(\ref{LVU}) for a $r\lambda^2F$ size pattern.\\

On Fig.(\ref{logt_inestables}), the average convergence time $T$ is plotted in logarithmic scale as a function of the initial number of unstable sites, for several initial type patterns satisfying $D_Y\geq D_L$ \footnote{There were generated 30 initial conditions for every particular pattern satisfying $D_Y\geq D_L$. The dimension error was always less than $\delta D=0.01$. In {\bf \ref{appendixA}}, we explain how the patterns were generated.}. For comparison, random initial conditions are also considered. From this perspective, structured  initial conditions really make the difference. Actually, they generally show a convergence time that is between 10 and 100 times higher than the random case. In turn, the less structured initial patterns ($\alpha=0$) are the closest ones to the random case, in terms of convergence time.  The convergence time is mostly increasing with $r$. In addition, note that the closer the size patterns ($Y$ or $L$) are to $\lambda^2F/2$, the closer is the system's response to the case of random initial conditions.\\  

On Fig.(\ref{logt}), convergence time is plotted as a function of the initial $D_Y$, for several fixed pattern fractions $r$'s. The convergence time $T$ is highly increasing with $D_Y$ for minimally structured $Y$-patterns, while the opposite approximately occurs for the maximally structured ones. In between these two regimes, $T$ follows a non-monotonous dependence with respect to $D_Y$.\\
For low dimensions, the higher initial $D_Y$, the harder for the system to stabilize because the individuals have more means to interact. The lower $D_Y$, the faster the system converges, mainly because individuals have more taboos ($M_{ijk}=0$) and tend to not interact between each other.\\
However, for high dimensions, the initial structure of the individuals somehow promotes convergence as $D_Y$ increases. In that case, there are many initial local structure $Y$-clusters for higher $D_Y$. This fact induces the system to reach local consensus sooner.\\
For intermediate dimensions, it is harder to say which of the two above scenarios prevails, increasing system unpredictability.
\end{section}

\begin{section}{The entropic formalism.}\label{an_entropic_formalism}
We suggest the above is in conection with the general complex system formalism developed in \cite{gaudiano}.
The article introduces the concept of {\it controllability} which in general refers to the possibility of driving a system from a given initial state to a final one by means of fields or constraints of any kind \cite{ogata}. In \cite{gaudiano} it is considered the notion that {\it what cannot be controlled is precisely what is ignored}. Thus, a maximum degree of uncontrollability could be given by system’s entropy (in a generalized sense), because the latter quantity is usually associated to {\it ignorance}.\\
In \cite{gaudiano}, a vast class of complex systems is studied assuming they present structures with hierarchical organization which in turn, is modelled in terms of fractal dimension ($D$). It is possible to define an entropy function $S(D)$ having many self-similar properties that ultimately define regimes of {\it uncontrollability}.\\

On the one hand, for the case of the Axelrod model of the present article, the unpredictability of the intermediate regime commented in the last section agrees with the uncontrollability ideas explored in \cite{gaudiano} (Fig.(\ref{logt})). 
On the other hand, from \cite{gaudiano}, the log number $S$ (entropy) of the pixel patterns (satisfying Eq.(\ref{LVU})) having fractal dimension $d$ is:
\begin{equation}\label{SN}
S(d)=\frac{F\lambda^2}{(2^{m-1})^N}H(2^{d-N})\frac{2^{md}-1}{2^d-1},\qquad{0\leq d\leq N}
\end{equation}
where $H(x)=-x\log_2(x)-(1-x)\log_2(1-x)$ is Shannon's binary entropy function. Fig.(\ref{SSpSpp}) shows that $S$ is a non-monotonous one hump function of $d$. As pointed out in \cite{gaudiano}, the zeros $D_k$ of $S$' $k$-th derivative  naturally define regimes which are characterized in terms of uncontrollability. In particular, the range $I_{21}$ defined as
\begin{equation}\label{D2DD1}
2.157...\equiv D_2\leq d\leq D_1\equiv 2.407...
\end{equation}
is suggested to be the most uncontrollable regime because of its high and increasing entropy (Fig.(\ref{SSpSpp})).\\
The article \cite{gaudiano} defines regime $I_{21}$ in general. Depending on the system under consideration, $I_{21}$ has been shown to correspond to a number of {\it out-of-control} situations: a kind of most divergent electoral scenario \cite{entropico}, a most segregationist regime \cite{schelling_entropico}, urban sprawl \cite{encarnacao}, out of control deforestation \cite{jinsun} and the most problematic public transport strike season \cite{lucca}.\\
For the present Axelrod model, it is remarkable how the convergence time behaves in a non-monotonous way within a range that coincides with the $I_{21}$ regime of Eq.(\ref{D2DD1}) (Fig.(\ref{logt})). As mentioned above, it implies uncontrollability. This fact will be visualized from another perspective, in the following section. 
\begin{figure}
\includegraphics[width=0.5\textwidth]{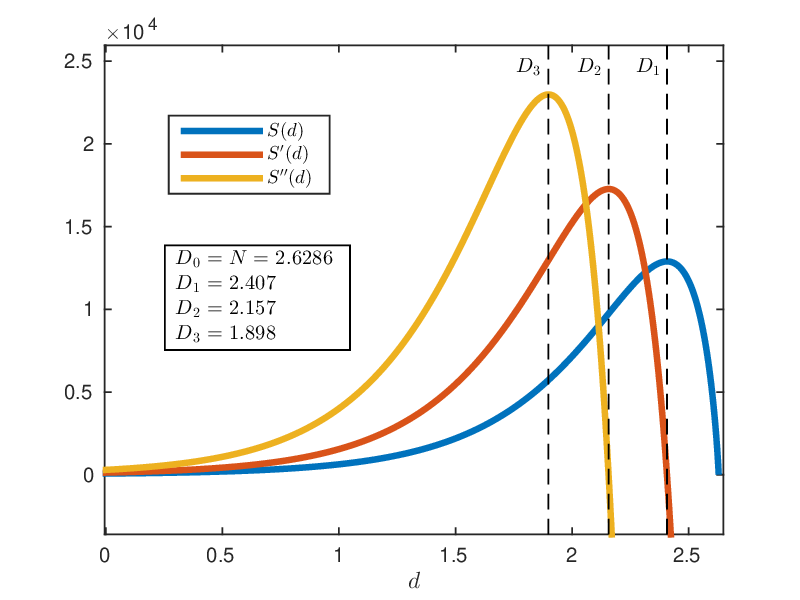}
\caption{Entropy function $S(d)$ and its first 2 derivatives.}
\label{SSpSpp}
\end{figure}
\end{section}

\begin{section}{Cultures.}\label{cultures} 
We say that an individual $(i,j)$ belongs to the {\it culture} $c=(c_1,...,c_F)\in\prod_{k=1}^F\{-1,0,1\}$ at time $t$ if
\begin{equation}\label{Mijk_ck}
M_{ijk}(t)=c_k \qquad{k=1,...,F.}
\end{equation}
This terminology is not just because of historical reasons \cite{axelrod}. It will be useful here because it gathers together individuals having the same attributes or opinions \cite{tangled}.\\  
Hierarchical organization properties can be naturally associated to every culture $c$. For example, once the system has converged, one can assess the fractal dimension $D(c)$ of the individuals $(i,j)$'s belonging to $c$. It tells us about the spatial (planar) extension of the individuals belonging to a same culture. Alternatively, given a culture $c$ we could explore the dimension $D_N(c)$ of the final total pattern of  active pixels $(i,j,k)$'s verifying Eq.(\ref{Mijk_ck})\footnote{note that $0\leq D(c)\leq 2$ and $0\leq D_N(c)\leq N$}. It may account for a kind of spatial-ideologic nature of cultures, visualizing each of them as a whole and not only in terms of the location of their constituents.\\

With an abuse of notation, $\sigma_{D_N(c)}$ ($\sigma_{D(c)}$) will denote the standard deviation of $D_N(c)$ ($D(c)$) varying $c$ \cite{sigmax}. As a function of the initial $D_Y$, these standard deviations are plotted on Fig.(\ref{sigmaDc_DYIII}) for {\it null taboo} cultures $c$'s ($c_k\neq 0$, $k=1,...,F$)
, which are the ones that prevail over the rest (because the inactive attributes are assumed time invariant). In this case, $D_N(c)$ and $D(c)$ are clearly related by
\begin{equation}\label{DNc}
D_N(c)=D(c)+N-2
\end{equation}
and consequently, their respective standard deviations coincide between each other despite of representing distinct culture aspects.\\

\begin{figure}
\includegraphics[width=0.5\textwidth]{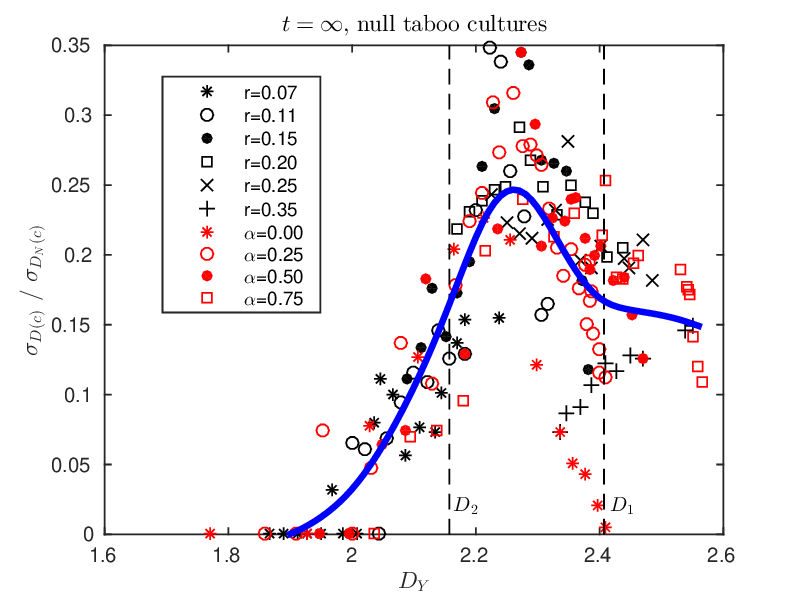}
\caption{As a function of the initial $D_Y$, these are the standard deviations at $t=\infty$ of $D(c)$ and $D_N(c)$ varying $c$ across null taboo cultures. 
Both standard deviations are coincident because of Eq.(\ref{DNc}).}
\label{sigmaDc_DYIII}
\end{figure}
However, the most important point concerning Fig.(\ref{sigmaDc_DYIII}) is that the variation of the fractal dimension of culture patterns maximizes on the range $I_{21}$. The standard deviations (\cite{sigmax}) of Fig.(\ref{amarillotabu0}) clearly points to the same direction, despite of corresponding to just the $Y$-part of the final null taboo culture patterns.\\

The above extends to cultures having a few taboos reasonably well (see Fig.(\ref{tabu1_2_012345678}) of {\bf \ref{appendixB}}), though they are not the ones that prevail over the rest at $t=\infty$. However, for those cases, the optimization of the standard deviation sometimes has not to do with bell-like distribution widenings but with the ocurrence of structure gaps and degeneration (Fig.(\ref{DY_DL_Dmin_histogramas})).\\
 All this is in very good agreement with the above commented uncontrollability regarding the range $I_{21}$. It reinforces the above commented conexion with the concept of {\it unpredictability} \cite{gaudiano} because of the diversification of the ultimate scenarios (both in extension, structure and content).
\begin{figure}
\centering
\includegraphics[width=0.5\textwidth]{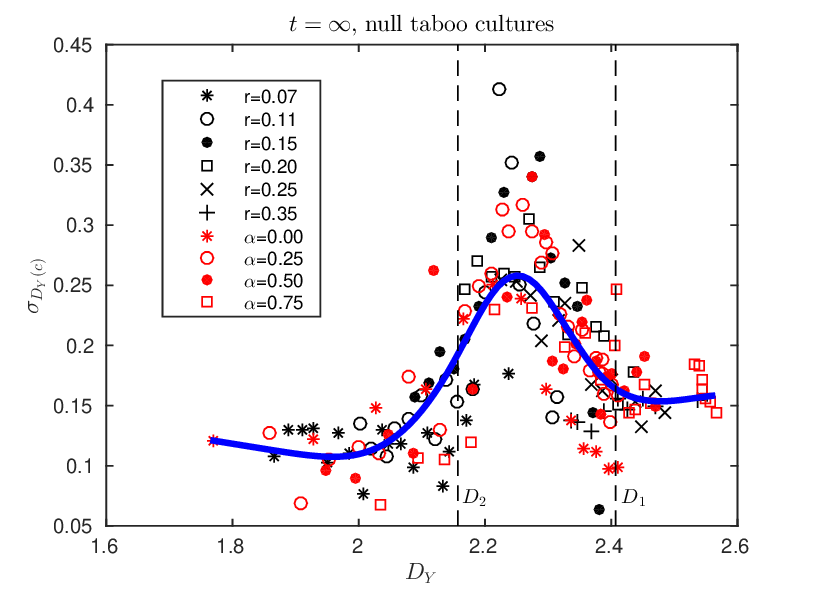}
\caption{As a function of the initial $D_Y$, these are the standard deviations of the dimensions $D_Y(c)$ corresponding to the $Y$-part of final 3-D patterns of null taboo cultures ($0\leq D_Y(c)\leq D_N(c)$).}
\label{amarillotabu0}
\end{figure}
\begin{figure}
\includegraphics[width=0.5\textwidth]{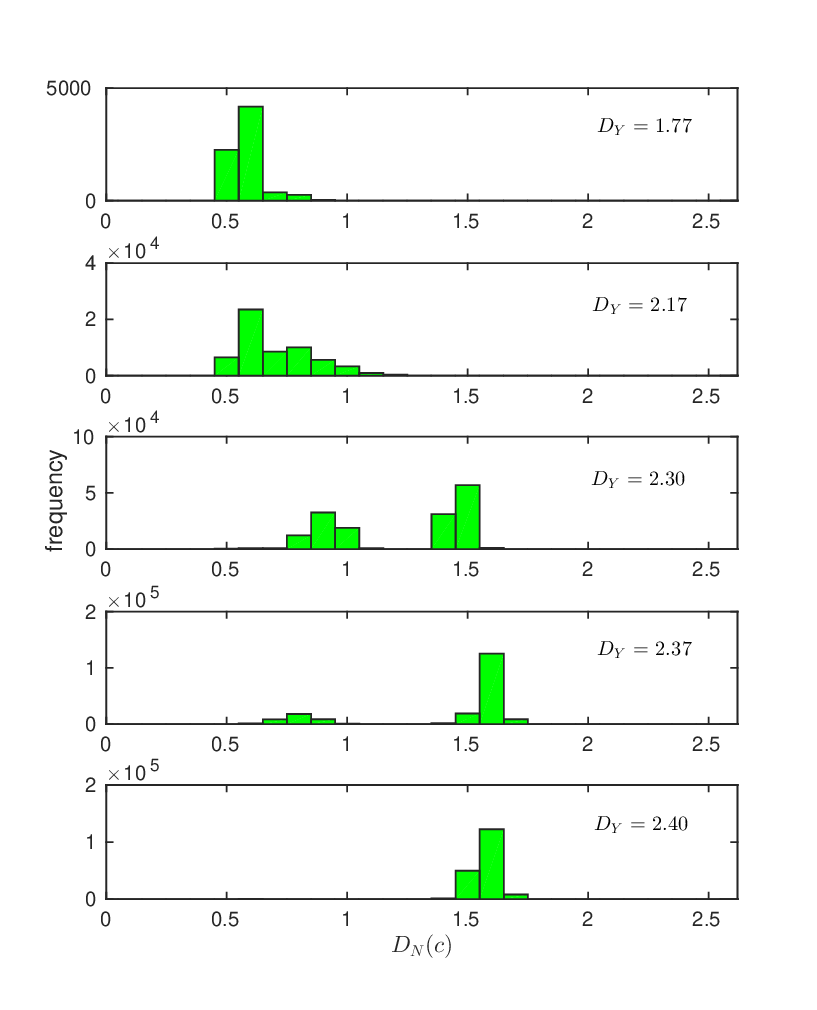}
\caption{Histograms of final pattern dimensions $D_N(c)$ of cultures having two taboos. At $t=0$, it is assumed that $D_Y=D_L=D_{min}(r)$ (from top to bottom: $r=0.05$, $0.15$, $0.30$, $0.40$ and $0.45$, respectively). Note that the standard deviation of $D_N(c)$ maximizes for $D_Y$ in the range $I_{21}$ (Eq.(\ref{D2DD1})).}
\label{DY_DL_Dmin_histogramas}
\end{figure}
\end{section}

\begin{section}{Quantization.}\label{quantization}

 
Despite Sociophysics quantization phenomena have been already observed and studied, for instance, in bounded confidence models (see e.g. \cite{deffuant}), we claim that Fig.(\ref{DY_DL_Dmin_histogramas}) unveils a quantization of different nature.\\
In order to understand the existence of the structure gaps of Fig.(\ref{DY_DL_Dmin_histogramas}), it is convenient to visualize them in a more general context. Given a culture $c$, its {\it taboo number} will be just the number of null attributes. For a moment, let us consider the case $r=0.35$ and $\alpha=0$ shown on Fig.(\ref{hist01}), which turns out to be somehow typical and consequently useful to illustrate the main ideas underlying the gaps. The figure shows an anomalous diffusion process in which the culture taboo numbers seem to play an important role (Fig. (\ref{hist01}, bottom)). The observed peaks can be roughly explained as follows.\\

\begin{figure}
\includegraphics[width=0.5\textwidth]{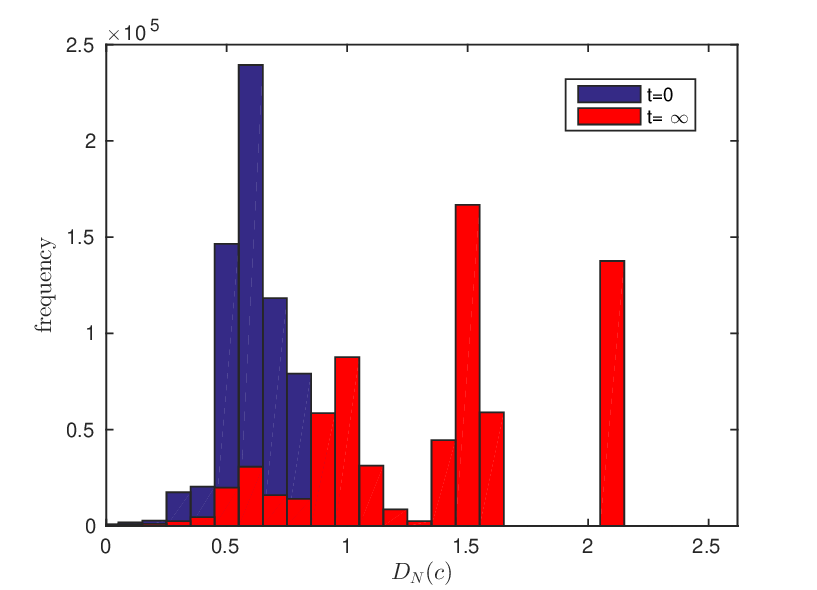}
\includegraphics[width=0.5\textwidth]{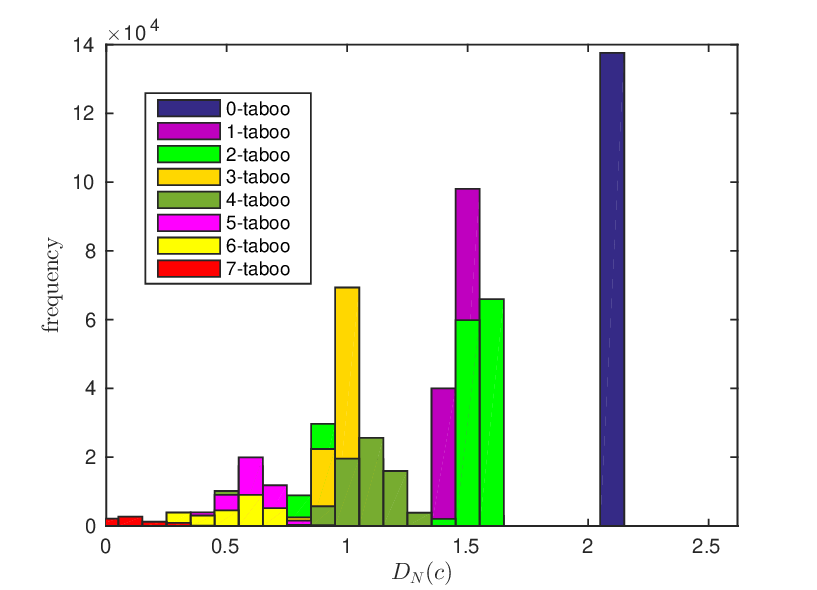}
\caption{Top: Initial and final distributions of the culture dimensions $D_N(c)$ for all simulations with $r=0.35$ and $\alpha=0$. Bottom: The same final histogram, but decomposing it by culture taboo numbers.}
\label{hist01}
\end{figure}

Let us define the $border$ of a prevailing culture $c$ (e.g. a null-taboo one) as the individuals $(i,j)$ belonging to it that have at least one adjacent (Section \ref{the_model}) site that does not belong to $c$. According to \cite{schelling_entropico}, it can be shown that the fractal dimension $D_\partial(c)$ of the borders 
satisfies
\begin{equation}\label{Dpartialc2DeltaD}
D_\partial(c)\lesssim 2-\Delta D
\end{equation}
where $\Delta D \simeq \langle D_{k}-D_{k+1}\rangle_k$ (see also \cite{gaudiano}). As stated in \cite{schelling_entropico}, Eq.(\ref{Dpartialc2DeltaD}) constitutes a purely geometrical fact. It is absolutely independent of the particular sociophysics model studied.\\

Moreover, Eq.(\ref{Dpartialc2DeltaD}) suggests that prevailing culture borders can be understood as lying on some space of dimensionality $2-\Delta D$. Analogously as in Eq.(\ref{SN}), an entropy function can be associated to that space. Consequently, if one supposes the most probable configuration for the borders, it may be assumed that
\begin{equation}\label{Dpartialc22DeltaD}
\langle D_\partial(c)\rangle\simeq 2-2\Delta D
\end{equation}   
which is the point of maximum of this new entropy function \cite{Snd}-\cite{gaudiano}. Thus, from Eq.(\ref{DNc}) we have
\begin{equation}\label{DNc2p13}
\langle D_N(c)\rangle\gtrsim N-2\Delta D\simeq 2.13
\end{equation}
by assuming the estimate $\Delta D\simeq 0.25$ (see \cite{Snd}). The above points out a correspondence between the first peak (from de right) of Fig.(\ref{hist01}) and the most prevailing culture average pattern.\\

When the system reaches equilibrium, we can consider succesive layers of individuals surrounding the set of sites of some null-taboo culture that prevails over the rest.  They belong to cultures $c$'s that have an average increasing taboo number. The first layer will be constituted by sites that are adjacent to the border of the prevailing culture. In principle, they may belong to cultures $c$'s that are equal to the prevailing one, except for the fact of having a few taboos (mainly 1 o 2, but not uniquely that). The second layer will be conformed by individuals that are adjacent to sites of the first layer. They belong to cultures $c$'s that are less similar to the null-taboo one and have generally higher taboo numbers, etc.\\
We conjecture that the above commented principles underlying Eqs.(\ref{Dpartialc2DeltaD}) and (\ref{Dpartialc22DeltaD}) somehow also work for the layers. Specifically, starting from Eq.(\ref{Dpartialc22DeltaD}), we could succesively subtract $2\Delta D$ in order to guess the dimensions of the layers. By doing so, we suggest that the most probable ({\it planar}) dimension of culture $c$ sites lying on the $i^{th}$ layer could be estimated as $2-2(i+1)\Delta D$. Under this supposition, it is reasonably easy to infer\footnote{note that the fractal dimension of a finite union of disjoint sets having dimensions $\delta_j$ ($j=1,2,3,..$) is given by $\max_{j}\delta_j$  \cite{falconer}}
 that {\it volumetric} dimension of the non-prevailing cultures should cluster around:
\begin{equation}\label{DNic}
\langle D_N(c)\rangle\simeq 2-2(i+1)\Delta D+\Delta N\qquad{i=1,2,3,...}
\end{equation}
where $\Delta N$ is an average perturbation accounting for the {\it z-axis} part of the culture 3-D patterns (like the term $N-2$ of  Eq.(\ref{DNc})). By inspection, it is not hard to assume $\Delta N\simeq 0.50$, at least for the first layers 
 ($i=1,2,3$). Under these assumptions, Eq.(\ref{DNic}) gives the values 1.50, 1.00 and 0.50, which approximately correspond to the dimension peaks of Fig.(\ref{hist01}).\\

There are some analogies to Quantum Physics here that are worthy to comment.\\
First, it is remarkable that for a single run, every culture $c$ will have a unique and precise value $D_N(c)$ (generally one of the peaks predicted by Eqs.(\ref{DNc2p13}) and (\ref{DNic})), despite of the degeneration of some culture taboo numbers (Figs.(\ref{DY_DL_Dmin_histogramas}) and (\ref{tabu1_2_012345678}, bottom)).\\
Second, according to Eq.(\ref{DNic}), the final cultures whose individuals conform patterns with null planar dimension will correspond to the peak $\langle D_N(c)\rangle\simeq \Delta N>0$.  Taking into account that this coincides to the peak at $t=0$ (Fig.(\ref{hist01}, top)), the minimally structured culture patterns somehow constitute a kind of {\it zero-point energy} state (in the absence of applied fields, etc.).\\ 

On Fig.(\ref{histogramas_alfa_r84_pesandoxvolumen}), the whole picture can be visualized. The numbers 0.50, 1.00 and 1.50 constitute a kind of {\it magic numbers} that are somehow present across mostly all histograms of the figure. At least for high $r$ or $\alpha$,  not only the locations of the peaks roughly coincide with the values predicted by Eq.(\ref{DNic}), but also the bound of Eq.(\ref{DNc2p13}) can also be recognized. It is worthy to note that these situations correspond to the cases in which the highest dimension of the final leading cultures is greater than 2. When the opposite occurs, we suggest that most of the cultures actually satisfy
\begin{equation}\label{DNc22DeltaD1p50}
D_N(c)\lesssim 2-2\Delta D\simeq 1.50
\end{equation}
(see {\bf \ref{appendixC}}). This estimate generally corresponds to low $r$ and $\alpha$ histograms of Fig.(\ref{histogramas_alfa_r84_pesandoxvolumen}) and implies a dimension gap between 1.50 and 2 for the highest possible culture dimensions. In this case, note the final culture dimensions also cluster around the minimum structure peak ($\Delta N$).\\

The existence of degeneration peaks now seems to be natural for some culture taboo numbers (Figs. (\ref{DY_DL_Dmin_histogramas}), (\ref{hist01}, bottom)) and (\ref{tabu1_2_012345678}, bottom). However, the most {\it pure state} culture patterns  correspond to the lowest taboo numbers. In contrast, the dimensions of the highest taboo number cultures gather arround $\Delta N$, generally remain invariant and correspond to cultures that almost never prevail over the rest.\\


Finally, note that for low $r$, the system almost does not evolve, becoming irrelevant the supposition of hierarchically organized initial configurations. The top histograms of Fig.(\ref{histogramas_alfa_r84_pesandoxvolumen}) show that dimension distributions are bell-shaped.\\  
In turn, it is remarkable that for every $\alpha$, the highest volumetric culture dimension tends to $N$ and its peak becomes the dominant one as $r$ increases. In addition, from the histograms, it is expected that the closer $r$ is to 1/2 (i.e. the sites mostly belong to non-taboo cultures), the less the initial structure conditions determine the evolution. In place of it, the system dynamics is basically a diffusion process as the classical Axelrod model, in which the space is often fulfilled by a single culture as $t\to\infty$ \cite{castellano_fortunato}.\\
All the above is seen in Fig.(\ref{logt_inestables}) since random and structured initial conditions produce almost equal convergence times at the limit of both low and high $r$'s.\\ 
\begin{figure*}
\includegraphics[width=1\textwidth]{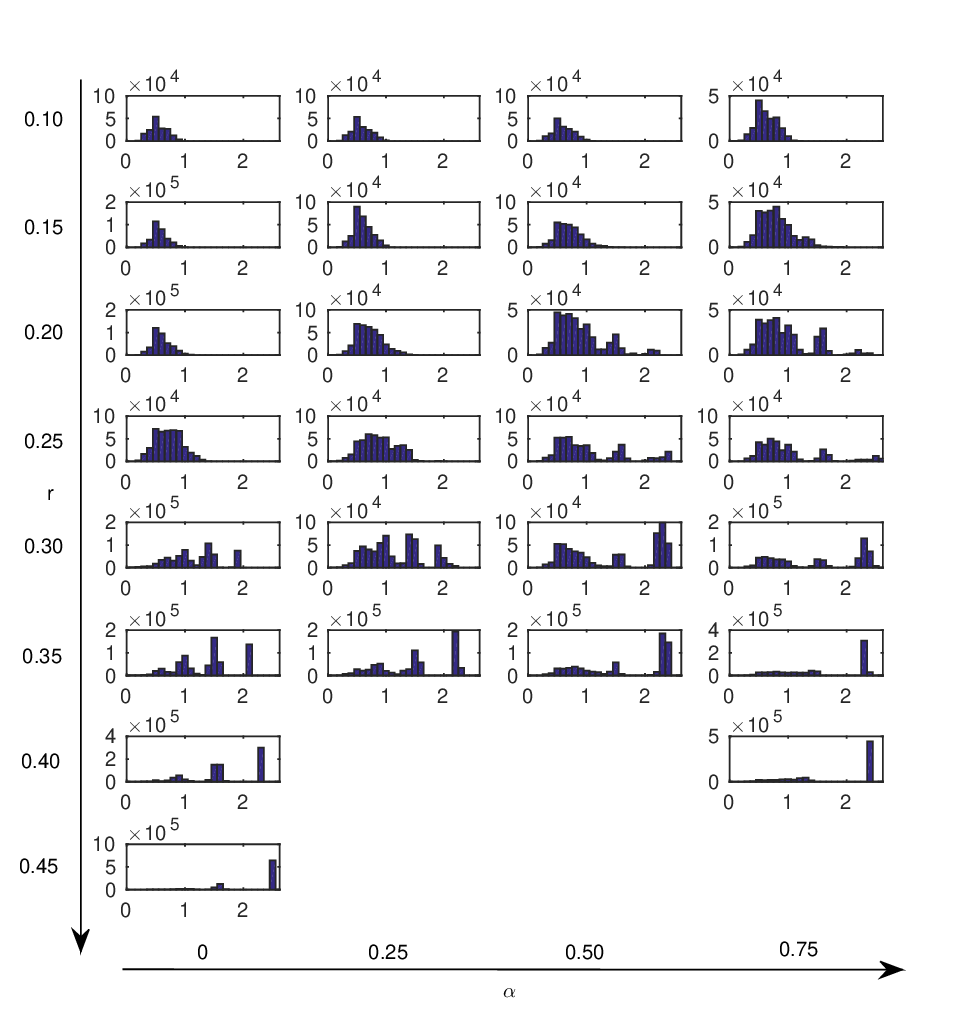}
\caption{Histograms of the dimensions $D_N(c)$'s at $t=\infty$ corresponding to non null cultures $c$ (i.e. $\max_k|c_k|=1$) varying $r$ and $\alpha$. There are some empty spaces (e.g. $r=0.45$ and $\alpha=0.25$) because of finite size effects.}
\label{histogramas_alfa_r84_pesandoxvolumen}
\end{figure*}
\end{section}

\begin{section}{Conclusion}\label{conclusion}
 Assuming hierarchically organized initial conditions in the Axelrod model produces outputs that notably differ from the random case. There are two limit cases that correspond to the Axelrod model with random initial conditions. For low $r$, interactions are scarse so the system almost does not evolve. For $r\to 1/2$, diffusion drives system dynamics and hierarchically organized initial conditions are not determinant.\\
 However,  when $r$ is not close to one of these limits, system response is strongly dependent on the structured initial conditions. These facts are remarkably depicted on both Figs. (\ref{logt_inestables}) and (\ref{histogramas_alfa_r84_pesandoxvolumen}).\\  

The description of the system in terms of the entropic regimes of \cite{gaudiano} has proven to be meaningful, pointing out a maximum unpredictability scenario ($I_{21}$), in plenty correspondence to the previous results of \cite{third_position}, \cite{entropico} and \cite{schelling_entropico}. Remarkably, these articles are about 2-D sociophysics models, and the present work shows that the ideas of \cite{gaudiano} can be extended to models of both higher and fractional dimensionality.\\

Structured initial conditions also allow to observe a resonance-like effect governed by entropy. When the entropy of the system is ranged in the uncontrollability regime $I_{21}$, the system optimizes its capacity to produce a more diverse set of ultimate scenarios. 
Outside that regime, the opposite happens and consequently, controllability increases.\\

In addition, studying Axelrod model under the formalism of \cite{gaudiano} provides an explanation to a system ultimate culture pattern quantization. This phenomenon is related to the culture taboo numbers. By using the entropic formalism, we were able to find a simple linear equation for it, which also points out the multidimensional nature of cultures. Besides, information about the model dynamics is implicitly contained in this quantization formula.\\
It is worthy to mention that this quantum effect does not occur when the system is initially close to random (minimally structured and small size).\\


 This article is not a closed work about the implications of \cite{gaudiano} on Axelrod model. Instead, we have explored perhaps the most significant aspects and for sure this will trigger future research. For instance, to study how to find out which are the initial cultures that become the  ultimate predominant ones and/or which cultures could be the ones the spontaneously emerge. The effect of increasing $F$ should also be an interesting issue \cite{futurework}.
\end{section}

\appendix
\section{Generation of fractal patterns.}\label{appendixA}
Let us imagine a pattern of (single) dimension $D$ comformed by a volume of $V$ pixels of the $\lambda\times\lambda\times F$ system tensor. The pattern can be naturally assumed as lying inside of a bigger tensor of size $\lambda\times\lambda\times\lambda$. Thus, according to the box counting method \cite{theiler}, when dividing the bigger tensor in cubes of side $\epsilon_j=2^j$ ($j=0,1,2,...,m=\log_2{\lambda/2}$), the number $N_j$ of cubes covering the pattern satisfies: 
\begin{equation}\label{box_counting}
\log N_j = -D \log\epsilon_j + \log V \qquad{j=0,1,2,...,m.}
\end{equation}

Now, in order to create a pattern lying inside of the $\lambda\times\lambda\times F$ tensor that consists of two separate parts having equal volume $V$ but different dimensions $D_L$ and $D_Y$, we will proceed as follows. First, two  vectors of integer random numbers $(N^{\beta}_1,...,N^{\beta}_m)$ ($\beta=L,Y$) are generated satisfying Eq.(\ref{box_counting}) for $V$ and $D=D_\beta\pm\delta D$ ($\beta=L,Y$), where $\delta D$ is a predetermined dimension error. Second, it is considered a random sequence of $2V$ letters consisting of $V$ $L$'s and $V$ $Y$'s. Third, $2V$ different and random pixels of the $\lambda\times\lambda\times F$ tensor will be successively chosen. The $i$-th pixel ($i=1,...,2V$) will be coloured yellow (light blue) if the $i$-th sequence letter is $Y$ ($L$). However, at each step, the number of $\epsilon_j$-side cubes covering the yellow (light blue) part of the being generated pattern (considered to be contained in the $\lambda\times\lambda\times\lambda$ bigger tensor) has not to exceed $N_j^Y$ ($N_j^L$), for $j=0,...,m$. If this cannot be accomplished or the final two coloured pattern does not satisfy the respective error requirements for $D_L$ and/or $D_Y$, the procedure starts again.\\

It was assumed that $\delta D=0.01$ throughout this work. Numerically, it is possible to observe this method is able to generate every required pattern, efficiently.\\

\begin{figure}
\includegraphics[width=0.5\textwidth]{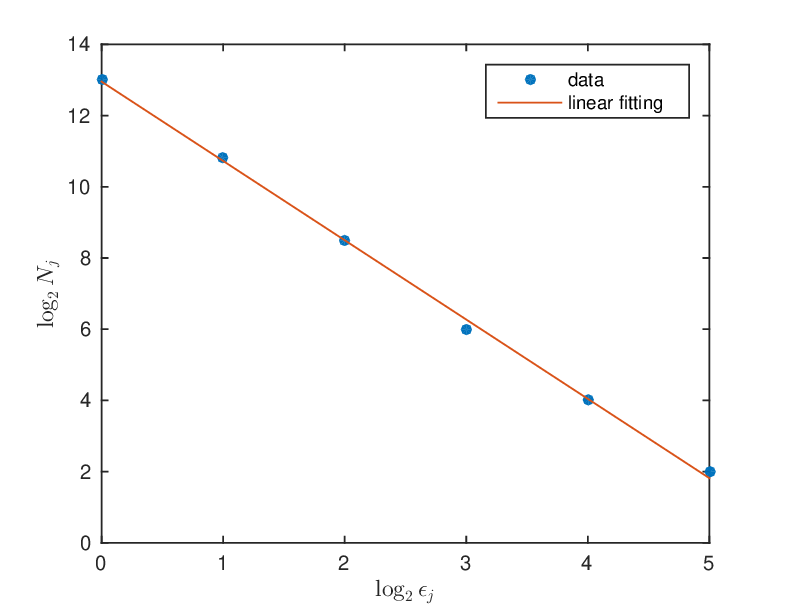}
\caption{The box-counting for the $L$-part top pattern of Fig.(\ref{T0T}). The numbers of cubes covering the pattern at each step of the box-counting are $N_0=8192$, $N_1=        1814$, $N_2=358$, $N_3=64$, $N_4=16$ and $N_5=4$. The absolute value of the slope of the linear fitting is the fractal dimension $D_L$, which is approximately 2.23.}
\label{example}
\end{figure}

It is worth to remark that the presented method generates hierarchically organized patterns, without supposing any further constraints. As an example, we reproduce on Fig.(\ref{example}), the box counting algorithm that computes the fractal dimension ($D_L$) of the $L$-part ($M_{ijk}=-1$) of the top pattern of Fig.(\ref{T0T}). It is assumed that $M_{ijk}=0$ for $F<k\leq \lambda$.

\section{Explanation of Equation (\ref{DNc22DeltaD1p50}).}\label{appendixC}
Let us suppose that the highest ultimate dimensions correspond to null taboo leading cultures $c$ satisfying $D_N(c)<2$. Then, from Eq.(\ref{DNc}) one has that $D(c)<4-N$, which somehow means that the planar culture patterns lie in spaces of dimensionalities that are lower than $4-N$. 
Thus, we conjecture that it is possible to replace (respectively) Eqs.(\ref{Dpartialc2DeltaD}) and (\ref{Dpartialc22DeltaD}) by $D_\partial(c)\lesssim 4-N-\Delta D$ and
$$\langle D_\partial(c)\rangle \lesssim 4-N-2\Delta D\approx 0.871...<1.$$
However, the latter implies that the corresponding culture borders are mostly constituted by disconnected points \cite{falconer}. Thus, it is not hard to infer that every culture planar pattern considered here should basically coincide to its own border. Thus, by averaging Eq.(\ref{DNc}) and introducing into it the above equation, it follows that the prevailing cultures satisfy that
$$\langle D_N(c)\rangle\approx \langle D_\partial(c)\rangle+N-2\lesssim 2-2\Delta D.$$
Consequently, Eq.(\ref{DNc22DeltaD1p50}) should hold for most of the cultures. 

\section{Some figures about $\sigma_{D_Y(c)}$.}\label{appendixB}
\begin{figure}
\includegraphics[width=0.5\textwidth]{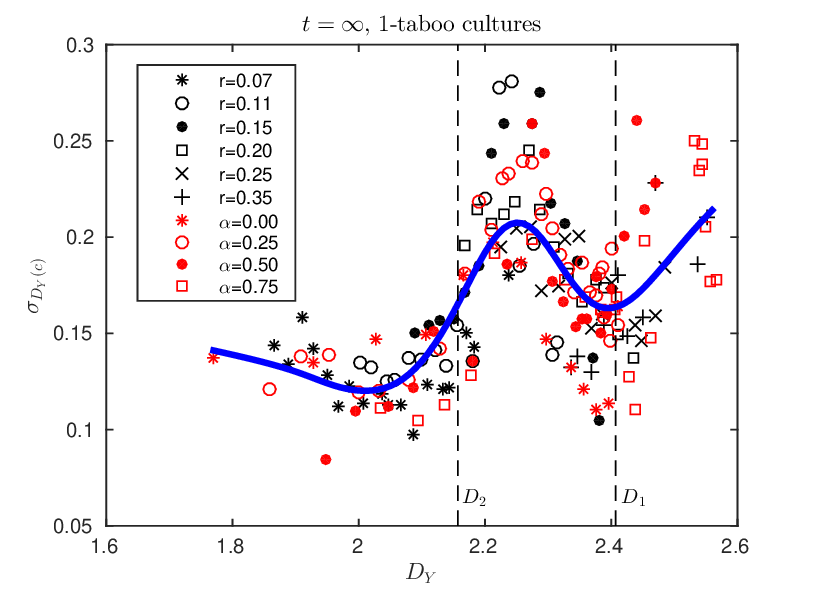}
\includegraphics[width=0.5\textwidth]{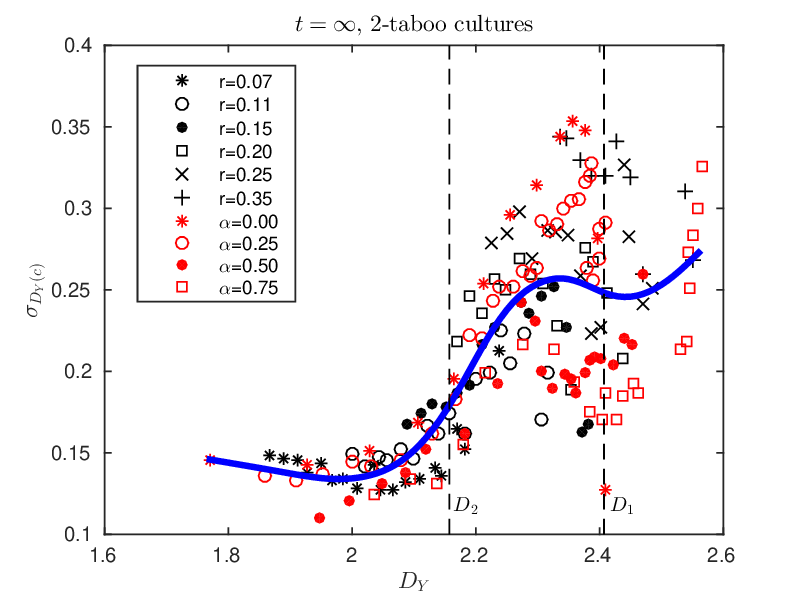}
\includegraphics[width=0.5\textwidth]{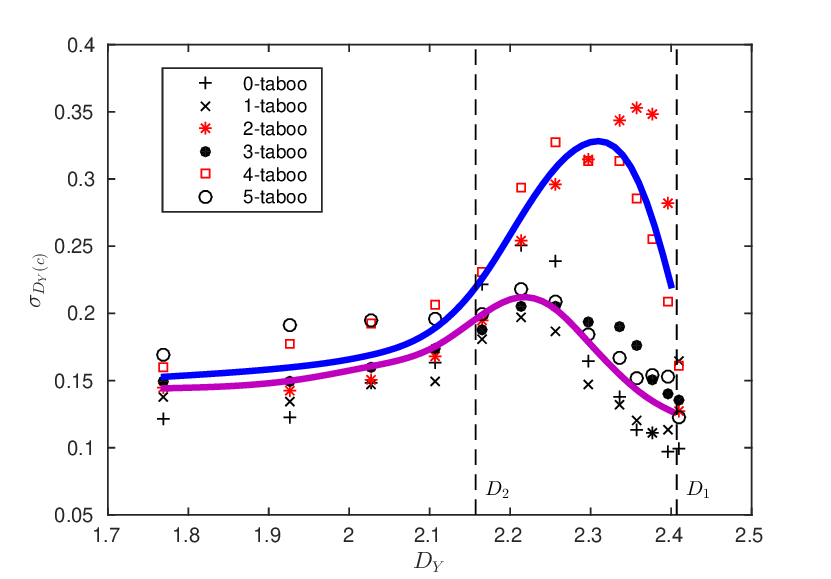}
\caption{Similar to Fig.(\ref{amarillotabu0}), but considering 1-taboo (top) and 2-taboo (center) cultures. Bottom: The case $D_Y=D_L=D_{min}(r)$ (i.e. $\alpha=0$) for cultures having only $j$ taboos ($j=0,...,5$).}
\label{tabu1_2_012345678}
\end{figure}

Fig. (\ref{tabu1_2_012345678}) was referenced in Section \ref{cultures}. It is about the standard deviation of the $Y-$part of the final culture patterns for three special cases. In every case, the respective $\sigma_{D_Y(c)}$ is maximized on the range $I_{21}$. Fig. (\ref{tabu1_2_012345678}, bottom) shows that the 2-taboo cultures reach higher standard deviation values, which correspond to the pair of peaks observed on Fig.(\ref{DY_DL_Dmin_histogramas}). 4-taboo culture standard deviations optimize in the same way.
\\
\\
{\bf Author Contribution Statement:} All authors contributed equally to the paper.\\


\begin{thebibliography}{9}
\bibitem{pumain} D. Pumain, {\it Hierarchy in Natural and Social Sciences} (Springer, Dordrecht, 2006).
\bibitem{encarnacao} S. Encarna\c{c}\~{a}o, M. Gaudiano, F.C. Santos, J.A. Tened\'orio and J.M. Pacheco (2012). Fractal Cartography of Urban Areas. Scientific Reports {\bf 2}, 527.
\bibitem{falconer} K.J. Falconer, {\it Fractal geometry: mathematical foundations and applications} (John Wiley \& Sons Inc, 2003).
\bibitem{jinsun} J. Sun, Z. Huang, Q. Zhen, J. Southworth and S. Perz (2014). Fractally Deforested Landscape: Pattern and Process in a Tri-national Amazon Frontier. Appl. Geogr. {\bf 52}, 204.
\bibitem{lucca} M.E. Gaudiano, C.M. Lucca and J.A. Revelli (2022). Entropic Analysis of Public Transport System Strikes. Advances in Complex Systems, Vol. 24, No. 6, 2250002. DOI: $10.1142/S0219525922500023$.
\bibitem{gaudiano} M.E. Gaudiano (2015). An Entropical Characterization for Complex Systems
Becoming out of Control. Physica A {\bf 440}, 185.
\bibitem{third_position} M.E. Gaudiano \& J.A. Revelli (2019). Spontaneous Emergence of a Third Position in an Opinion Formation Model. Physica A {\bf 521}, 501.
\bibitem{entropico} M.E. Gaudiano \& J.A. Revelli (2021). Entropical Analysis of an Opinion Formation Model Presenting a Spontaneous Third Position Emergence. Eur. Phys. J. B 94:89. https://doi.org/10.1140/epjb/s10051-021-00098-8
\bibitem{schelling_entropico} M.E. Gaudiano \& J.A. Revelli (2021). On the role of structured initial conditions in the Schelling model. Physica A, Vol. 587.
\bibitem{axelrod} R. Axelrod (1997). The dissemination of culture: A model with local convergence and global
polarization. J. Conflict Resolut. 41, 203-226.
\bibitem{axelrod_media}L. R. Peres and J. F. Fontanari (2011). The media effect in Axelrod's model explained. EPL, Vol. 96, 38004. DOI: $10.1209/0295-5075/96/38004$.
\bibitem{tangled} H. Rajpal, F. Rosas and H. Jensen (2019). Tangled Worldview Model of Opinion
Dynamics. Front. Phys. 7:163. DOI: $10.3389/fphy.2019.00163$.
\bibitem{castellano_fortunato} C. Castellano, S. Fortunato and V. Loreto (2009). Statistical physics of social dynamics. Rev. Mod. Phys. 81, 591.
\bibitem{morgan} B. Morgan (1984). Social Geography, Spatial Structure and Social Structure. GeoJournal 9, no. 3: 301-10. $http://www.jstor.org/stable/41143395$.
\bibitem{theiler} Theiler, J., Estimating fractal dimension, J. Opt. Soc. Am. A 7 (1990) 1055-1073.
\bibitem{ogata} K. Ogata, {\it Modern Control Engineering (3rd ed.)} (Prentice-Hall, Upper Saddle River, NJ, 1997). ISBN 978-0-13-227307-7.
\bibitem{sigmax} If $X(c,s)$ and $W(c,s)$ are the $s^{th}$ simulation dimension and size of a given (2D or 3D) culture $c$ pattern, then
$$\sigma_{X(c)}^2\doteq \sum_{s,c} p_{cs}(X(c,s)-\bar{d})^2,$$
 where $\bar{d}=\sum_{s,c}p_{cs}X(c,s)$ and $p_{cs}=W(c,s)/\sum_{s,c}W(c,s)$. Note that if the patterns under consideration involve cultures having a common number of taboos, both $\sigma_{D(c)}$ and $\sigma_{D_N(c)}$ can be indistinctly computed with $W(c,s)=$ No. of  $(i,j)$'s  belonging to $c$ at $t=\infty$ (area) or  $W(c,s)=$ No. of  $(i,j,k)$'s  such that $(i,j)$ belongs to $c$  at $t=\infty$  and $|M_{ijk}(\infty)|=1$ (volume). This will be approximately the general idea for the rest of the article (Figs. (\ref{amarillotabu0})-(\ref{histogramas_alfa_r84_pesandoxvolumen}) and (\ref{tabu1_2_012345678}) look basically the same if weighted by area or volume). It reinforces the multidimensional nature of cultures.\\
\bibitem{deffuant} G. Deffuant, D. Neau, F. Amblard and G. Weisbuch (2000), Adv. Complex Syst. 3, 87.
\bibitem{Snd} According to \cite{gaudiano}, if $S_n(d)$ is the total log number of fractal patterns having dimension $d$ in a space of arbitrary dimension $n$ ($0\leq d\leq n$), the zeros of its $k$-th derivative with respect to $d$ will be approximately given by $d_k=n-k\Delta D$ ($k=0,1,2,...$), being $\Delta D$ almost independent of $n$. For this reason, $\Delta D$ can be indistinctly estimated with the zeros $D_k$'s of the succesive derivative of $S(d)$ (Eq.(\ref{SN})): $\Delta D\approx D_1-D_2\approx 0.25$ (Eq.(\ref{D2DD1})). Moreover, the right side of Eq.(\ref{Dpartialc22DeltaD}) is the point of maximum $d_1$ of $S_{2-\Delta D}(d)$, i.e. $d_1=(2-\Delta D)-\Delta D=2-2\Delta D$. 
\bibitem{futurework} M.E. Gaudiano and J.A. Revelli. Paper in preparation.
\end{thebibliography}
\end{document}